\documentclass[12pt,a4paper]{article}

\usepackage{epsfig}
\usepackage{amsmath}
\usepackage{verbatim}
\usepackage{psfrag}
\usepackage{bm}

\numberwithin{equation}{section}

\begin{document}

\begin{flushright}
{\tt MAN/HEP/2007/11}\\
{\tt SPIN-07/20, ITP-UU-07/30} 
\end{flushright}

\vskip .5cm

\begin{center}

{\Large\bf Baryogenesis from the amplification of vacuum fluctuations
during inflation}

\vskip 1cm

Bj\"orn~Garbrecht$^1$ and Tomislav Prokopec$^2$\\

\vskip .4cm

{\it
${}^1$School of Physics \& Astronomy,
The University of~Manchester,\\
Oxford~Road,
Manchester M13~9PL, United Kingdom\\
$^2$Institute for Theoretical Physics (ITF) \& Spinoza Institute,
Utrecht University,\\
Leuvenlaan 4, Postbus 80.195, 3508 TD Utrecht, The Netherlands
}

\end{center}

\vskip .5cm

\begin{abstract}
We propose that the baryon asymmetry of the Universe may originate from
the amplification of quantum fluctuations of a light complex scalar field
during inflation. CP-violation is sourced by complex mass terms, which
are smaller than the Hubble rate, as well as non-standard
kinetic terms. We find that, when assuming 60 e-folds of inflation, an
asymmetry in accordance with observation can result for models where
the energy scale of inflation is of the order of $10^{16}{\rm GeV}$. 
Lower scales may be achieved when assuming substantially larger
amounts of e-folds.
\end{abstract}

\section{Introduction}

It was pointed out by Sakharov that the matter-antimatter asymmetry of
the Universe does not need to be put in as an initial condition
set up at the Big Bang, but can be explained within models of
particle physics, even if the initial asymmetry is assumed
to be zero~\cite{Sakharov:1967dj}. This is an important and celebrated
observation, and in fact the failure of the Standard Model
to explain the baryon number of the Universe (BAU) is one of the
chief motivations for considering its extensions.
The inflationary paradigm further augments the urge for an
explanation of the BAU from particle physics, since due to the
increase of its volume by at least a factor of about
$10^{70}$, the Universe is essentially void of particles at the end of
inflation. Any possible pre-existing asymmetry would have been 
diluted away.

Loopholes to this argument appear to exist. Recently, a model within
an extended, CP-violating Brans-Dicke model
and CP violation in the kinetic term has been
proposed~\cite{vanderPost:2006nd}. The authors have shown that 
an initial classical lump of scalar charge can be amplified and stretched 
to large scales during inflation. After inflation the charge is then 
converted into a baryon asymmetry. 

While Ref.~\cite{vanderPost:2006nd}
deals with amplification of classical charge density, here we show that,
analogously to cosmological perturbations, a similar scalar field model
can be used to amplify quantum charge density fluctuations during inflation.
In this paper, we point out that indeed, a negative mass term is not necessary
to generate a sufficient BAU. For a complex scalar field with CP-violating
masses and kinetic terms, the growth of the modes due to the expanding
background can lead to a sufficient BAU, provided the physical
mass eigenvalues for the scalar degrees of freedom do not exceed the Hubble
expansion rate. The mechanism of particle production is the same as for
the generation of density perturbations from inflation, supplemented however
with CP-violation. 

 For simplicity we perform our analysis in de Sitter space. We show 
that charged quantum fluctuations are generically amplified to
a charge density that is on super-Hubble wavelengths independent on scale
(scale invariant spectrum). Moreover, when integrated out, 
at the end of inflation one obtains a scalar charge density $q_\phi$ 
of the form (see Eq.~(\ref{q:dSinvariant})),
\begin{equation}
q_\phi=|\omega \mu^2| H \sin \theta_{\rm CP}
\left[
-\frac{3}{2\pi^2}\frac{H^2 \dot H}{m^4 - |\mu|^4}
+\frac{1}{8\pi^2}\frac{1}{\varrho^2 -|\omega|^2}
\right]\,,
\label{q:dSinvariant:Intro}
\end{equation}
where $\mu$ and $m$ are the CP odd and even masses, $\rho$ and $\omega$
are the (dimensionless) couplings of the CP even and odd kinetic terms,
respectively ({\it cf.} Eq.~(\ref{L:non_stan:phi})) 
and $\dot H=dH/dt$ is the rate of change 
of the Hubble parameter $H$. As expected, 
according to the result~(\ref{q:dSinvariant:Intro}) 
the charge vanishes either when CP violation vanishes
 (which is the case when any of the parameters $\omega$, $\mu$ 
or $\theta_{\rm CP}$ vanishes) or when $H$ vanishes
 (because the Universe's expansion is the source for the amplification 
of charged vacuum fluctuations). The result~(\ref{q:dSinvariant:Intro})
allows us to establish a relation between inflationary observables
and baryogenesis, which we discuss in some detail
 in section~\ref{section:BAU}.

 Conserved currents usually give a unique prescription to define charges.
In turn, when there is no exactly conserved current, the definition of
a charge is in general ambiguous. In practice, this is most often not a
problem due to the existence of hierarchies in physical models. For example,
consider standard leptogenesis. Within the Standard Model, the baryon minus
lepton number $B-L$ is conserved, but when Majorana neutrinos are added,
it is not. Nonetheless, since the Majorana neutrinos are usually either
chosen to be much heavier than Standard Model particles or very weakly coupled
to these,
it is of course reasonable to define $B-L$ from the conserved global charge in
the Standard Model, even though it is violated in the full theory.

 In section~\ref{section:current}, we take the view that the scalar field,
which carries the initial charge asymmetry, decays into much lighter fermions,
which carry $B-L$ charge. This leads us to a current which is conserved in the
fermionic sector but violated for the scalar field, analogous to the
above example. In section~\ref{section:realfield} we quantize
the scalar field and make use of this to
calculate the current at late times during inflation
in section~\ref{section:scalarcurrent}. Possible particle physics models
as realizations of our mechanism are sketched in
section~\ref{section:decay}, and a quantitative discussion of the BAU
including the relation to the scale of inflation is the content
of~\ref{section:BAU}.

\section{CP violation and charged current
}
\label{section:current}

In this section, we introduce a scalar lagrangian which
is endowed with CP-violation,
such that a charge asymmetry can result from vacuum fluctuations.
In order to obtain a meaningful notion of a charge number, we also
define the coupling of the scalar  to fermions. This leads us to a
current that
is approximately conserved when the scalar field decays
into the fermions. The calculation of this current is the subject of
the subsequent sections.

The most general quadratic lagrangian for a single complex scalar field,
which is minimally coupled to curvature, is given by
\begin{equation}
{\cal L}_\phi = \sqrt{-g}\left[
           \varrho g^{\mu\nu}(\partial_\mu \phi) \partial_\nu \phi^*
   + \frac{\omega}{2} g^{\mu\nu}\partial_\mu\phi\partial_\nu \phi
   + \frac{\omega^*}{2} g^{\mu\nu}\partial_\mu\phi^*\partial_\nu \phi^*
   - m^2 \phi\phi^* 
   - \frac{\mu^2}{2}\phi^2 - \frac{{\mu^*}^2}{2}{\phi^*}^2 
  \right]\,.
\qquad
\label{L:non_stan:phi}
\end{equation}
We assume an inflationary background and use conformal coordinates,
such that the metric is
$g_{\mu\nu}=a^2 \eta_{\mu\nu}$ with the scale factor
\begin{equation}
\label{scale:factor}
a=-\frac1{H\eta}\,,
\end{equation}
where $\eta<0$ denotes the conformal time and $H$ the Hubble rate.
The deviation of inflationary expansion from
an exact de Sitter phase can be incorporated by allowing for a weak time
dependence of $H$. A rigorous treatment of the charge production 
in quasi de Sitter spaces can be performed by making use 
of the quasi de Sitter scalar propagator~\cite{JanssenProkopec:2007}.
We furthermore introduce a Dirac fermion coupling to the scalar field as
\begin{eqnarray}
{\cal L}_\psi &=& \sqrt{-g}\left[
           \overline\psi_L{\rm i}\partial\!\!\!/\psi_L
           + \overline\psi_R{\rm i}\partial\!\!\!/\psi_R
           - f\left(\overline\psi_R \phi \psi_L +{\rm h.c.}\right)
            \right]
\,,
\label{L:non_stan:psi}
\end{eqnarray}
where $f$ is a real Yukawa coupling. The full lagrangian density is then
given by
\begin{eqnarray}
{\cal L}&=& {\cal L}_\phi+{\cal L}_\psi\,.
\label{L:non_stan:sum}
\end{eqnarray}
It is of course possible to redefine $\phi$ in such a way that the
lagrangian~(\ref{L:non_stan:phi}) has a canonical kinetic term.
In section~\ref{section:realfield}, we perform such a redefinition
explicitly.
In that case, the Yukawa couplings in~(\ref{L:non_stan:psi}) get more
complicated. For the definition of a current, which is conserved
during the decays to fermions, we however find
the present field definition more useful.

Starting point for the definition of the current is a
${\rm U}(1)$-charge, under which
\begin{eqnarray}
\label{U1}
\psi_R &\to&{\rm e}^{-{\rm i}\beta}\psi_R\,, \nonumber\\
\psi_L &\to&{\rm e}^{-{\rm i}(\alpha+\beta)}\psi_L\,, \nonumber\\
\phi &\to& {\rm e}^{{\rm i}\alpha}\phi\,.
\end{eqnarray}
These transformation properties are chosen in such a way that the Yukawa terms
are singlets, {\it i.e.} during the decay of a $\phi$-particle to a pair
of fermions, the ${\rm U}(1)$-charge remains conserved. Up to a normalization,
this requirement defines the above symmetry and current in a unique way.
If after the complete decay of $\phi$ a certain charge
is stored in the $\psi_L$, we are interested in how to
express this charge as a function in terms of $\phi$ before the decay.

Next, we recall some facts about symmetry currents. Let
${\cal L}_\varphi$ be a lagrangian and
$\varphi$ a scalar field. Performing an infinitesimal transformation
$\varphi\to \varphi +\varepsilon \Delta \varphi$ has the effect 
${\cal L}_\varphi\to{\cal L}_\varphi+\varepsilon \Delta {\cal L}_\varphi$
on the lagrangian, where
\begin{equation}
\Delta{\cal L}_\varphi
=\frac{\partial {\cal L}_\varphi}{\partial \varphi}\Delta\varphi
+\left(\frac{\partial {\cal L}_\varphi}{\partial(\partial_\mu \varphi)}\right)
\partial_\mu\Delta\varphi
=
\partial_\mu\left(
\frac{\partial {\cal L}_\varphi}{\partial(\partial_\mu \varphi)}
\Delta\varphi\right)
+\left[
\frac{\partial {\cal L}_\varphi}{\partial \varphi}-
\partial_\mu \frac{\partial {\cal L}_\varphi}{\partial(\partial_\mu \varphi)}
\right]\Delta\varphi
\,,
\end{equation}
and the last term vanishes by virtue of the Euler-Lagrange equation.
If $\Delta {\cal L}_\varphi$ is zero or it is possible to write it as 
a total divergence $\Delta {\cal L}_\varphi= \partial_\mu {\cal J}^\mu$,
we can define the conserved current
\begin{equation}
j_\varphi^\mu
=\frac{\partial {\cal L}_\varphi}{\partial(\partial_\mu \varphi)}\Delta\varphi
-{\cal J}^\mu\,,
\end{equation}
which satisfies $\partial_\mu j_\varphi^\mu=0$.

For the symmetry~(\ref{U1}), we use
\begin{equation}
j_\phi^\mu
=\frac{\partial \cal L_\phi}{\partial(\partial_\mu \phi)}\Delta\phi\,,
\end{equation}
where $\Delta\phi={\rm i}\phi$ (taking in~(\ref{U1}) an infinitesimal
$\alpha=\varepsilon$), such that we find
\begin{eqnarray}
 j^\mu_\phi&=&\sqrt{g}g^{\mu\nu}\left\{{\rm i} \varrho \left[
(\partial_\nu \phi^*)\phi - \phi^* (\partial_\nu \phi)
\right]
+\left[
{\rm i} \omega (\partial_\nu \phi)\phi +{\rm c.c.}
\right]\right\}
\,.
\label{currents}
\end{eqnarray}
In this case however, $\Delta \cal L$ is non-zero and
can neither be expressed in terms of a
total divergence, which means that the symmetry~(\ref{U1}) is violated.
Explicitly,
\begin{equation}
\label{DL:violating}
\Delta {\cal L}= \sqrt{-g}\left\{{\rm i} \omega g^{\mu\nu}
\partial_\mu\phi \partial_\nu \phi
-{\rm i}\mu^2 \phi^2
+{\rm h.c.}
\right\}\,.
\end{equation}
This terms originates from ${\cal L}_\phi$ only and has no contribution
from ${\cal L}_\psi$, according to our requirement that the decay of the scalar
to fermions and the fermionic sector itself should respect the current.
Of course, the total current
$j_{\phi,\psi}^\mu = j^\mu_\phi+j^\mu_\psi$ also
has the fermionic contribution
\begin{eqnarray}
j^\mu_\psi=\sqrt{-g} \bar \psi_L \Gamma^\mu \psi_L
\,,
\end{eqnarray}
where $\Gamma^\mu$ denotes the $\gamma$-matrices
contracted with the vierbein linking curved space with a local
Minkowskian coordinate system. For a conformal background
$g_{\mu\nu}=a^2\eta_{\mu\nu}$,
as assumed in this paper, we have $\Gamma^\mu=(1/a)\gamma^\mu$

The ${\rm U}(1)$-violation during inflation as expressed
by~(\ref{DL:violating})
can be the source for successful baryogenesis.
On the other hand, we have to require that after inflation, the once generated
charge is not erased. It is therefore important that the field
$\phi$ does not thermalize and that it quickly decays into the fermions. 
Under the conditions
\begin{equation}
\label{smallCPviolation}
f^2 m /(8\pi) \stackrel{>}{{}_\sim} \mu\;\; \textnormal{and}\;\;
\varrho  \stackrel{>}{{}_\sim} \omega\,,
\end{equation}
the decay rate into  fermions is
\begin{equation}
\Gamma_{\phi\to{\bar\psi \psi}}\approx f^2\sqrt{ m^2/\varrho}/(8\pi)\,,
\end{equation}
whereas the rate of
${\rm U}(1)$-violation is of the order of 
\begin{equation}
\partial_0j^0/(j^0)=O(\mu/\sqrt\varrho,m \sqrt{\omega/\varrho}),
\end{equation}
such that the requirement
\begin{equation}
\label{fastdecay}
\partial_0j^0/(j^0)\stackrel{<}{{}_\sim}\Gamma_{\phi\to{\bar\psi \psi}}
\end{equation}
is satisfied.

\section{Real field decomposition}
\label{section:realfield}

In this section, we quantize the scalar field $\phi$ as defined by
the lagrangian~(\ref{L:non_stan:phi}). In order to do so, we decompose
$\phi$ into canonically normalized real degrees of freedom mixing through
a real symmetric mass-square matrix $M^2$.

For the complex scalar field $\phi$, it is always possible to go 
to the real field basis
\begin{equation}
 \phi  = \alpha \phi_+ + i\beta \phi_-
\,,\quad  
 \phi^\dagger  = \alpha^* \phi_+ - i\beta^* \phi_-
\,,\qquad
\left(\phi_+=\frac{\beta^*\phi+\beta\phi^\dagger}{2\Re[\alpha\beta^*]}
\,,
      \phi_-=\frac{\alpha^*\phi-\alpha\phi^\dagger}{2i\Re[\alpha\beta^*]}
\right)\,.
\label{real decomposition}
\end{equation}
Inserting this into the lagrangian~(\ref{L:non_stan:phi})
and demanding that the kinetic terms are canonically normalized
we obtain
\begin{eqnarray}
 - i \varrho \big(\alpha\beta^*-\alpha^*\beta\big)
 + i \omega\alpha\beta - i (\omega\alpha\beta)^* &=& 0\,,
\label{condition:1}
\\
 \varrho |\alpha|^2 + \frac{1}{2}\omega\alpha^2
                    + \frac{1}{2}(\omega\alpha^2)^* &=& \frac12\,,
\label{condition:2}
\\
 \varrho |\beta|^2 - \frac{1}{2}\omega\beta^2
                   - \frac{1}{2}(\omega\beta^2)^* &=& \frac12\,.
\label{condition:3}
\end{eqnarray}
With these substitutions the lagrangian~(\ref{L:non_stan:phi}) becomes
\begin{equation}
 {\cal L}_\phi = \sqrt{-g}\left[
     \frac12 \sum_{\pm}g^{\mu\nu}(\partial_\mu \phi_\pm) \partial_\nu \phi_\pm
   -\frac12 m_+^2 \phi_+^2
   -\frac12 m_-^2 \phi_-^2
   - m_{+-}^2 \phi_+\phi_-
  \right]
\,,
\label{L:non_stan:phi:3}
\end{equation}
where the mass-squares are 
\begin{eqnarray}
   m_+^2 &=& 2|\alpha|^2\left[
                           m^2 + |\mu|^2 \cos(2\theta_\mu+2\theta_\alpha)
                      \right]\,,
\label{m2+}\\
   m_-^2 &=& 2|\beta|^2\left[
                           m^2 - |\mu|^2 \cos(2\theta_\mu+2\theta_\beta)
                      \right]\,,
\label{m2-}\\
   m_{+-}^2 &=& 2|\alpha||\beta|\left[
                           m^2\sin(\theta_\alpha-\theta_\beta) 
                         - |\mu|^2 \sin(2\theta_\mu+\theta_\alpha+\theta_\beta)
                      \right]
\label{m2+-}
\label{L:non_stan:masses}
\end{eqnarray}
and where we defined
\begin{equation}
 \omega = |\omega| {\rm e}^{i\theta_\omega}
\,,\qquad
 \alpha = |\alpha| {\rm e}^{i\theta_\alpha}
\,,\qquad
 \beta = |\beta| {\rm e}^{i\theta_\beta}
\,.
\label{omega-alpha-beta}
\end{equation}

 Eqs.~(\ref{condition:1}--\ref{condition:3})
represent two real and one imaginary condition for two complex numbers,
which means that $\alpha$ and $\beta$ are {\it not} completely specified.  
 The conditions~(\ref{condition:1}--\ref{condition:3}) can be recast as
\begin{eqnarray}
 \varrho \sin(\theta_-)
  - |\omega|\sin(\theta_+)
            &=& 0\,,
\label{condition:1b}
\\
 \varrho |\alpha|^2 
+ \frac{1}{2}|\omega|\,|\alpha|^2\cos(\theta_++\theta_-)
                            &=& \frac12\,,
\label{condition:2b}
\\
 \varrho |\beta|^2 
+ \frac{1}{2}|\omega|\,|\beta|^2\cos(\theta_+-\theta_-)
                            &=& \frac12\,,
\label{condition:3b}
\end{eqnarray}
where we introduced the angles
\begin{equation}
  \theta_+ = \theta_\omega+\theta_\alpha+\theta_\beta
\,,\qquad
  \theta_- = \theta_\alpha-\theta_\beta
\,.
\label{theta+-}
\end{equation}
>From the first equation~(\ref{condition:1b}) we get
\begin{equation}
 \sin(\theta_-) = \frac{|\omega|}{\varrho}\sin(\theta_+)\,,
\label{condition:1c}
\end{equation}
while the latter two equations~(\ref{condition:2b}--\ref{condition:3b}) imply
\begin{eqnarray}
\cos(\theta_+) \cos(\theta_-) 
  &=& \frac{1}{4|\omega|} \left(\frac{1}{|\alpha|^2}-\frac{1}{|\beta|^2}\right)\,,
\label{condition:2c}
\\
\sin(\theta_+) \sin(\theta_-) 
  &=& \frac{\varrho}{|\omega|} 
     - \frac{1}{4|\omega|}\left(\frac{1}{|\alpha|^2}+\frac{1}{|\beta|^2}\right)\,.
\label{condition:3c}
\end{eqnarray}
 From~(\ref{condition:3c}) and~(\ref{condition:1c}) we then find
\begin{eqnarray}
\sin^2(\theta_-)
 &=& 1- \frac{1}{4\varrho}\left(\frac{1}{|\alpha|^2}+\frac{1}{|\beta|^2}\right)
\,.
\label{condition:3d}
\end{eqnarray}
Finally, Eq.~(\ref{condition:2c})
 can be solved to give the following relation between
 $|\alpha|$ and $|\beta|$,
\begin{equation}
  |\alpha|^2 + |\beta|^2 =  \frac{\varrho}{\varrho^2 - |\omega|^2}
\,.
\label{condition:2d}
\end{equation}

 As promised the solution is not unique. Indeed,
 there is a one parameter family of solutions. 
 A particularly convenient choice is
\begin{equation}
  |\alpha|^2 = |\beta|^2 = \frac12 \frac{\varrho}{\varrho^2 - |\omega|^2}
\,,
\label{condition:2e}
\end{equation}
for which we then have
\begin{equation}
 \sin(\theta_+) = 1
\,,\quad  
 \cos(\theta_+) = 0
\,,\quad  
 \theta_+ = \frac{\pi}{2}
\,,\qquad  
 \sin(\theta_-) = \frac{|\omega|}{\varrho}
\,,\quad  
 \cos(\theta_-) = \sqrt{1-\frac{|\omega|^2}{\varrho^2}}
\,.
\label{conditions:angles}
\end{equation}
 With this choice the mass-squares~(\ref{m2+}--\ref{m2+-}) become
\begin{eqnarray}
   m_+^2 &=& \frac{\varrho}{\varrho^2-|\omega|^2}\left[
                         m^2 + |\mu|^2 \cos(\theta_{\rm CP}+\theta_++\theta_-)
                      \right]
\nonumber\\
      &=& \frac{\varrho}{\varrho^2-|\omega|^2}
         \left[
        m^2
        - |\mu|^2 \left(\frac{|\omega|}{\varrho}\cos(\theta_{\rm CP})
                    +\sqrt{1-\frac{|\omega|^2}{\varrho^2}}\sin(\theta_{\rm CP})
                  \right)
         \right]\,,
\label{m2+:2}\\
   m_-^2 &=& \frac{\varrho}{\varrho^2-|\omega|^2}\left[
                          m^2 - |\mu|^2 \cos(\theta_{\rm CP}+\theta_+-\theta_-)
                      \right]
\nonumber\\
      &=& \frac{\varrho}{\varrho^2-|\omega|^2}
         \left[
        m^2
        - |\mu|^2 \left(\frac{|\omega|}{\varrho}\cos(\theta_{\rm CP})
                    -\sqrt{1-\frac{|\omega|^2}{\varrho^2}}\sin(\theta_{\rm CP})
                  \right)
         \right]\,,
\label{m2-:2}\\
   m_{+-}^2 &=& \frac{\varrho}{\varrho^2-|\omega|^2}\left[
                           m^2\sin(\theta_-) 
                         - |\mu|^2 \sin(\theta_{\rm CP}+\theta_+)
                      \right]
\nonumber\\
      &=& \frac{\varrho}{\varrho^2-|\omega|^2}
         \left[
        m^2\frac{|\omega|}{\varrho}
        - |\mu|^2\cos(\theta_{\rm CP})
         \right]
\,,
\label{m2+-:2}
\end{eqnarray}
where we defined the CP angle 
\begin{equation}
 \theta_{\rm CP} = 2\theta_\mu - \theta_\omega
\,.
\label{CP angle}
\end{equation}


%
%
%
%

The lagrangian~(\ref{L:non_stan:phi:3}) 
implies the following field equations:
\begin{eqnarray}
  \frac{1}{a^2}\left(a^2 \phi_+^\prime\right)^\prime 
     + (-\partial_i^2 + a^2 m_+^2)\phi_+ + a^2 m_{+-}^2 \phi_- &=& 0\,,
\nonumber\\
  \frac{1}{a^2}\left(a^2 \phi_-^\prime\right)^\prime 
     + (-\partial_i^2 + a^2 m_-^2)\phi_- + a^2 m_{+-}^2 \phi_+ &=& 0\,,
\label{eom:+-}
\end{eqnarray}
where a prime denotes a derivative with respect to $\eta$.
When introducing the mass-square matrix
\begin{equation}
 M^2 \equiv \left(\begin{array}{cc}
                      m_+^2 & m_{+-}^2 \cr
                      m_{+-}^2 & m_-^2 \cr
                  \end{array}
            \right)\,,
\label{M2:def}
\end{equation}
these can be compactly written as
\begin{equation}
  \frac{1}{a^2}\left(a^2 \Phi^\prime\right)^\prime 
     + (-\partial_i^2 + a^2 M^2)\Phi = 0\,,
\label{eom:Phi}  
\end{equation}
where we defined 
\begin{equation}
 \Phi = \left(\begin{array}{c}
                 \phi_+ \cr \phi_-
              \end{array}
        \right)
\,.
\label{Phi:def}
\end{equation}
The conformally rescaled form of Eq.~(\ref{eom:Phi}) is
\begin{equation}
 \left(\partial_\eta^2 - \partial_i^2 + a^2 M^2 
      - \frac{a^{\prime\prime}}{a}\right)(a\Phi) = 0
\,.
\label{eom:Phi:cf}  
\end{equation}

 We now perform the canonical quantization by imposing the commutator
\begin{equation}
 \left[\hat\phi_a(\vec x,\eta),\hat \pi_b(\vec x^{\,\prime},\eta)\right]
   = {\rm i} \delta_{ab}\delta^3(\vec x- \vec x^{\,\prime})
\,\qquad (a,b=\pm,\; \hbar=1)
\,
\label{canonical quantization}
\end{equation}
and the canonical momentum equals
\begin{equation}
  \hat\pi_a = \frac{\partial {\cal L}_\phi}{\partial\partial_\eta \phi_a}
        = a^2 \hat\phi_a^\prime \,,\qquad (a=\pm)
\,,
\end{equation}
or equivalently 
\begin{equation}
  \hat\Pi = a^2 \hat\Phi^\prime 
\,,
\label{canonical momentum:2}
\end{equation}
with 
\begin{equation}
 \hat\Phi = \left(\begin{array}{c}\hat\phi_+ \cr\hat\phi_-
                   \end{array}
                    \right)
\,,\qquad \hat\Pi = \left(\begin{array}{c}\hat\pi_+ \cr\hat\pi_-
                          \end{array}
                    \right)
\,.
\label{canonical momentum:3}
\end{equation}

 Since the system~(\ref{eom:+-})
 we want to solve is linear and the mass matrix~(\ref{M2:def})
is real symmetric,
the field can be expressed in terms of the annihilation and creation
operators as
\begin{equation}
  \phi_a(x) = \sum_{b=\pm}
       \int \frac{d^3k}{(2\pi)^3} {\rm e}^{{\rm i}\vec k\cdot \vec x}
           \left[
                 \phi_{k,ab}(\eta)\hat a_{\vec k,b}
               +  \phi^*_{k,ab}(\eta)\hat a^\dagger_{-\vec k,b}
           \right] 
\,,\qquad (a,b=\pm, k=\|\vec k\|)
\,,
\label{field decomposition}
\end{equation}
or in a matrix form, 
\begin{equation}
  \Phi(x) = \int \frac{d^3k}{(2\pi)^3} {\rm e}^{{\rm i}\vec k\cdot \vec x}
           \left[
                 \Phi_{k}(\eta)\cdot\hat A_{\vec k}
               +   \Phi^*_{k}(\eta) \cdot \hat A^\dagger_{-\vec k}
           \right]
\,,
\label{field decomposition:b}
\end{equation}
where 
\begin{equation}
  \Phi_k = \left(\begin{array}{cc}
                  \phi_{k,++} & \phi_{k,+-} \cr
                  \phi_{k,-+} & \phi_{k,--} \cr
                 \end{array}
           \right) 
\,,\qquad 
  \hat A_{\vec k} = \left(\begin{array}{c}
                  \hat a_{\vec k,+} \cr
                  \hat a_{\vec k,-} \cr
                 \end{array}
           \right) 
\,.
\end{equation}
Note that the symmetry of $M^2$ also implies that if $\Phi_k(\eta)$
and $\Phi_k^\prime(\eta)$
are symmetric for a particular $\eta$, they are symmetric at all
times. In particular, we impose the symmetry of $\Phi_k$ and $\Phi_k^\prime$
as initial
conditions and, in the following, make frequently use of the symmetry property
$\Phi_k^*(\eta)=\Phi^\dagger_k(\eta)$.

 Upon inserting~(\ref{field decomposition:b}) into~(\ref{eom:Phi:cf}) we get 
for the mode equation
\begin{equation}
 \left(\partial_\eta^2 +k^2 + a^2 M^2 
      - \frac{a^{\prime\prime}}{a}\right)(a\Phi_k(\eta)) = 0
\,.
\label{eom:Phi_k:cf}  
\end{equation}
 This equation can be solved in de Sitter space with the scale
factor~(\ref{scale:factor}).
 The two fundamental solutions can be written as
\begin{equation}
  \Phi_k(\eta) 
       = \frac{1}{a}\sqrt{-\frac{\pi\eta}{4}}{\rm e}^{{\rm i}(\pi/2)(\nu-1/2)}
                    H_\nu^{(1)}(-k\eta)
\,,\quad \Phi_k^\dagger(\eta)
\qquad  \left(\nu^2 = \frac{9}{4} - \frac{M^2}{H^2}\right)
\,.
\label{dS:mode function}
\end{equation}
This is a matrix generalization of the standard Bunch-Davies vacuum solution
for de Sitter space~\cite{Bunch:1978yq}.
Since the mass matrix is time-independent, 
the Hankel functions in~(\ref{dS:mode function}) can be ascribed
a precise meaning. Also, since there is only one matrix 
in~(\ref{dS:mode function}), the ordering of terms in 
Eq.~(\ref{dS:mode function}) is irrelevant. 
The general mode functions are then a linear combination of the fundamental 
solutions~(\ref{dS:mode function}),
\begin{equation}
\Xi = \gamma_1\cdot\Phi_k(\eta) + \gamma_2\cdot\Phi_k(\eta)^\dagger
\,,\qquad
\Xi^\dagger = \Phi_k^\dagger(\eta)\cdot\gamma_1 + \Phi_k(\eta)\cdot\gamma_2
\,,
\label{dS:general mode function}
\end{equation}
where $\gamma_1$ and $\gamma_2$ are constant $2\times 2$ matrices. 
Since $\nu$ is a constant matrix, it is easily checked
that the Wronskian of the fundamental solutions~(\ref{dS:mode function}) 
is the simple generalization of the single field case, 
\begin{equation}
  W[\Phi_k(\eta),\Phi^\dagger_k(\eta)]
     =  \Phi_k(\eta)\cdot{\Phi^\dagger_k}^\prime(\eta)
        - \Phi_k^\prime(\eta)\cdot\Phi^\dagger_k(\eta)
     = \frac{\rm i}{a^2}
\,.
\label{Wronskian:matrix}
\end{equation}
Requiring that the Wronskian of the general mode
 function~(\ref{dS:general mode function}) be unchanged leads to the following 
condition on $\gamma_1$ and $\gamma_2$,
\begin{equation}
 \gamma_1\cdot \gamma_1^\dagger - \gamma_2\cdot \gamma_2^\dagger = \mathbf{1}
\,.
\label{condition on gammas}
\end{equation}
In this work we shall take the initial vacuum state
corresponding to zero particles as $\eta\to-\infty$, in which case
we have $\gamma_1 = {\rm diag}(1,1)$ and $\gamma_2={\rm diag}(0,0)$.

\section{Scalar current}
\label{section:scalarcurrent}

In this section, we apply the scalar field decomposition and
quantization to calculate the scalar charge generated during inflation.
The comoving scalar current~(\ref{currents}) in conformal coordinates
is
\begin{equation}
 \hat{ j_\phi}^\mu =
a^4
\left\{ {\rm i} \varrho 
    \left[ (\partial^\mu \hat\phi^\dagger)\hat\phi
          - \hat\phi^\dagger (\partial^\mu \hat\phi)
    \right]
  +  {\rm i} \omega (\partial^\mu \hat\phi)\hat\phi
         - {\rm i} \omega^* (\partial^\mu \hat\phi^\dagger)\hat\phi^\dagger
\right\}
\,,
\label{current:phi}
\end{equation}
which, when expressed in terms of the real fields, becomes
\begin{eqnarray}
 \hat{ j_\phi}^\mu 
 \!\!\!&=&\!\!\! a^4\Big\{
\left[{\rm i} \omega\alpha^2-{\rm i} (\omega\alpha^2)^*\right] 
     \hat\phi_+\partial^\mu \hat\phi_+
   - \left[{\rm i} \omega\beta^2-{\rm i} (\omega\beta^2)^*\right] 
     \hat\phi_-(\partial^\mu \hat\phi_-)
\label{current:phi:2}
\\
 &+&\!\!\! \left[ -\varrho(\alpha\beta^*+\alpha^*\beta)
           \!-\! \omega\alpha\beta\!-\!(\omega\alpha\beta)^*\right] 
                 \hat\phi_-\partial^\mu \hat\phi_+
   + \left[\varrho(\alpha\beta^*+\alpha^*\beta)
           \!-\! \omega\alpha\beta\!-\!(\omega\alpha\beta)^*\right] 
                 \hat\phi_+\partial^\mu \hat\phi_-
\Big\}
\,.
\nonumber
\end{eqnarray}
%
%
%
With the choice~(\ref{conditions:angles}) this reduces to 
\begin{equation}
 \hat{ j_\phi}^\mu 
= a^4 \left\{-\frac{|\omega|}{\sqrt{\varrho^2-|\omega|^2}}
      \left[\hat\phi_+\partial^\mu \hat\phi_+
         - \hat\phi_-\partial^\mu \hat\phi_-
      \right]
 + \frac{\varrho}{\sqrt{\varrho^2-|\omega|^2}}
   \left[
         \hat\phi_+\partial^\mu \hat\phi_-
      -  \hat\phi_-\partial^\mu \hat\phi_+
   \right] 
\right\}
\,.\quad
\label{current:phi:4}
\end{equation}

 We now use the solution~(\ref{field decomposition:b}) 
and~(\ref{dS:mode function})
to calculate the expectation value of the current~(\ref{current:phi:4})
 in the vacuum $\hat a_{\vec k,b}|0\rangle = 0$. Noting
$\partial_\mu=(\partial_\eta,\partial/\partial x_i)$, we find
\begin{eqnarray} 
 {j_\phi}^\mu 
 & \equiv& \langle 0| \hat { j_\phi}^\mu |0\rangle 
\label{current:phi:5}
\\
  &=& -a^2 \frac{|\omega|}{\sqrt{\varrho^2-|\omega|^2}}
      \int \frac{d^3k}{(2\pi)^3} \left[
 \left(\Phi_k(\eta)\cdot (\partial_\eta,ik_i) \Phi^\dagger_k(\eta)\right)_{++}
 - \left(\Phi_k(\eta)\cdot(\partial_\eta,ik_i)\Phi^\dagger_k(\eta)\right) _{--}
      \right]
\nonumber\\
 &&+ a^2 \frac{\varrho}{\sqrt{\varrho^2-|\omega|^2}}
   \int \frac{d^3k}{(2\pi)^3} 
     \left[
  \left(\Phi_k(\eta)\cdot (\partial_\eta,ik_i) \Phi^\dagger_k(\eta)\right)_{+-}
 -\left(\Phi_k(\eta)\cdot(\partial_\eta,ik_i) \Phi^\dagger_k(\eta)\right)_{-+}
   \right]
\,,
\nonumber
\end{eqnarray} 
where we made use of 
\begin{equation}
  \langle 0| \hat \Phi\cdot \partial_\mu \hat\Phi |0\rangle 
   =  \int \frac{d^3k}{(2\pi)^3}
          \Phi_k(\eta)\cdot \partial_\mu \Phi^\dagger_k(\eta)\,.
\label{exp.v. for current}
\end{equation}
The spatial part of the current~(\ref{current:phi:5}) vanishes
when integrated over the momentum, so that we need to calculate only 
the 0th component.
Moreover, since the mass matrix is symmetric, so is $\Phi$, and hence 
$\Phi^\dagger = \Phi^*$. This allows one to rewrite the time component 
of Eq.~(\ref{exp.v. for current}) as
\begin{equation}
          \Phi_k(\eta)\cdot \partial_\eta \Phi^\dagger_k(\eta)
        = \frac{i}{2a^2}
    + \frac12\partial_\eta\left\{\Phi_k(\eta)\cdot\Phi^*_k(\eta)\right\} 
\,,
\label{exp.v. for current:time}
\end{equation}
where we made use of the Wronskian~(\ref{Wronskian:matrix}).

To get the physical charge density $q_\phi$, we recall that
$j_\phi^0$ corresponds to the comoving charge. Therefore,
one ought to multiply~(\ref{current:phi:5}) by $1/a^3$:
\begin{eqnarray} 
 q_\phi &=& \frac{ { j_\phi}^0}{a^3}
  = -\frac{1}{a}\frac{|\omega|}{\sqrt{\varrho^2-|\omega|^2}}
      \int \frac{d^3k}{(2\pi)^3} \left[
 \left(\Phi_k(\eta)\cdot \partial_\eta \Phi^\dagger_k(\eta)\right)_{++}
 - \left(\Phi_k(\eta)\cdot\partial_\eta\Phi^\dagger_k(\eta)\right) _{--}
      \right]
\nonumber\\
 &&+ \frac{1}{a}\frac{\varrho}{\sqrt{\varrho^2-|\omega|^2}}
   \int \frac{d^3k}{(2\pi)^3} 
     \left[
  \left(\Phi_k(\eta)\cdot \partial_\eta \Phi^\dagger_k(\eta)\right)_{+-}
 -\left(\Phi_k(\eta)\cdot\partial_\eta \Phi^\dagger_k(\eta)\right)_{-+}
   \right]
\,.
\label{current:phi:6}
\end{eqnarray} 
%
When the symmetry of the mass matrix
and hence of $\Phi_k(\eta)$
under transposition is taken account of, 
one sees that there is no contribution
 from the second line in~(\ref{current:phi:6}).

 Since the Wronskian in~(\ref{exp.v. for current:time}) does not contribute
to the charge~(\ref{current:phi:6}), it can be written
as a derivative with respect to the physical time $t$,  which is related to
the conformal time by $dt=ad\eta$,
\begin{eqnarray} 
 q_\phi &=& -\frac12\frac{|\omega|}{\sqrt{\varrho^2-|\omega|^2}}\partial_t
      \int \frac{d^3k}{(2\pi)^3} \left[
 \left|\Phi_k(\eta)\right|^2_{++} - \left|\Phi_k(\eta)\right|^2_{--}
      \right]
\nonumber\\
 &&+ \frac12\frac{\varrho}{\sqrt{\varrho^2-|\omega|^2}}
   \partial_t\int \frac{d^3k}{(2\pi)^3} 
     \left[
  \left|\Phi_k(\eta)\right|^2_{+-} -\left|\Phi_k(\eta)\right|^2_{-+}
   \right]
\,.
\label{current:phi:7}
\end{eqnarray} 
The above integrals are well known from
the correlator ${\rm i}\Delta(x,x^\prime)$ for the scalar field, which
can be represented in terms of the mode sum
\begin{equation}
{\rm i}\Delta(x,x^\prime)
=\langle0|\Phi(x^\prime)\Phi^\dagger(x)|0\rangle
=\int \frac{d^3 k}{(2\pi)^3}
\Phi_k(\eta){\rm e}^{\rm i \mathbf{k}\cdot\mathbf{x}}
\Phi_k^*(\eta^\prime){\rm e}^{-\rm i \mathbf{k}\cdot\mathbf{x^\prime}}\,.
\end{equation}
We recall that the Chernikov-Tagirov solution for
this propagator in terms of hypergeometric functions is~\cite{Chernikov:1968zm}
\begin{eqnarray}
{\rm i}\Delta(x,x^\prime)\!\!\!&=&\!\!\!
\frac{\Gamma(\frac32+\nu)
    \Gamma(\frac32-\nu)}{(4\pi)^2
   }H^2\;
{}_2F_1\Big(\frac32+\nu,\frac32-\nu;2;1-\frac{y}{4}\Big)
\label{Delta:Exact}
\,,
\end{eqnarray}
and can be expanded as~\cite{Prokopec:2003tm,Garbrecht:2006jm}
\begin{eqnarray}
{\rm i}\Delta(x,x^\prime)\!\!\!&=&\!\!\!
\frac{H^2}{4\pi^2}\frac 1y
+\frac{H^2}{16\pi^2}\sum\limits_{n=0}^\infty
\frac{\Gamma\left(\frac 32 + \nu +n\right)\Gamma\left(\frac 32 - \nu +n\right)}
{\Gamma\left(\frac 12 +\nu \right)\Gamma(\frac 12 -\nu)}
\left(\frac y4 \right)^n
\\
&&\!\!\!\times
\left[
\log\frac y4
+ \psi\left(\frac 32 +\nu +n\right)+ \psi\left(\frac 32 -\nu +n\right)
-\psi\left( 1+n \right) - \psi\left(2+n\right)
\right]\,,
\nonumber
\end{eqnarray}
where
\begin{equation}
y=\frac{-(\eta-\eta^\prime)^2+\mathbf{x}^2}{\eta\eta^\prime}\,.
\end{equation}
When furthermore expanding in $M^2/H^2$, we find
\begin{eqnarray}
\label{Delta:Expanded}
{\rm i}\Delta(x,x^\prime)\!\!\!&=&\!\!\!
\frac{H^2}{4\pi^2}\frac 1y +\frac{3H^4}{8\pi^2M^2}
+\left(\frac{\log2}{4\pi^2}-\frac{7}{24\pi^2}-\frac{\log y}{8\pi^2}\right)H^2
\nonumber\\
&&\!\!\!
+\left(
\frac{\log y}{16\pi^2} - \frac{\log2}{8\pi^2}- \frac 1{216\pi^2} +O(y)
\right)M^2+O\left(\frac{M^4}{H^2}\right)\,.
\end{eqnarray}
Applying this to compute the charge~(\ref{current:phi:7}), we remember that
only the $++$ and $--$ components lead to a
non-vanishing result. The off-diagonal terms cancel due to the
symmetry of $M^2$ and consequently of ${\rm i}\Delta$.
>From Eq.~(\ref{exp.v. for current:time}) it follows that 
the charge density is obtained by symmetrizing the time derivative as
follows, 
\begin{eqnarray}
\label{q:symmetric}
q_\phi=-\frac{|\omega|}{\varrho^2-|\omega|^2}
\frac12(\partial_t+\partial_{t^\prime})
\left[
\langle 0|
\Phi_+(x^\prime)\Phi_+(x)
|0 \rangle
-
\langle 0|
\Phi_-(x^\prime)\Phi_-(x)
|0 \rangle
\right]\Big|_{x=x^\prime}\,.
\end{eqnarray}
The term $(\partial_t+\partial_{t^\prime})(1/y)|_{x=x^\prime}$
is singular, but these contributions cancel in~(\ref{q:symmetric}),
since the coefficient of $1/y$ in~(\ref{Delta:Expanded}) is independent
of $M^2$. The leading contributions to $q_\phi$ therefore stem from
the terms $\propto 1/M^2$ and $\propto \log y \,M^2$. Noting
$(\partial_t+\partial_{t^\prime})\log y|_{x=x^\prime}=H$, we find
\begin{equation}
\label{q:dSinvariant}
q_\phi=|\omega \mu^2| H \sin \theta_{\rm CP}
\left[
-\frac{3}{2\pi^2}\frac{H^2 \dot H}{m^4 - |\mu|^4}
+\frac{1}{8\pi^2}\frac{1}{\varrho^2 -|\omega|^2}
\right]\,,
\end{equation}
where the dot stands for a derivative with respect to $t$.
The above expression is the main result of this paper. 
In appendix~\ref{appendix:IR}, we additionally present a derivation
of the charge spectrum on super-Hubble (infrared) scales, while
appendix~\ref{appendix:UV} contains the corresponding sub-Hubble
(ultraviolet) expansion.

Let us now summarize the assumptions and approximations
we have used to derive this expression for
the scalar charge $q_\phi$.
First, the expansion~(\ref{Delta:Expanded})
is valid when $|M^2|\stackrel{<}{{}_\sim} H^2$.
To check the applicability of the
result~(\ref{q:dSinvariant}), $H^2$ has to be compared to
the eigenvalues of the mass-squared matrix~(\ref{M2:def}),
\begin{equation}
M^2_\pm=\frac{\varrho}{\varrho^2-|\omega|^2}
\left[
m^2-\frac{|\omega \mu^2|}\varrho \cos\theta_{\rm CP}
\pm\sqrt{
|\mu|^4\left(1-\frac{|\omega|^2}{\varrho^2}\right)
+m^4\frac{|\omega|^2}{\varrho^2}
-2\frac{|\omega\mu^2|}{\varrho^2}m^2 \cos\theta_{\rm CP}
}
\right]\,.
\label{Mpm}
\end{equation}
Second, in order to have approximate charge conservation after inflation
until the decay of $\phi$, relation~(\ref{fastdecay}) needs to hold.
Taking account of these, we note that a resonant CP violation
by setting $m\simeq|\mu|$ or $\varrho \simeq |\omega|$ cannot be achieved
and that the condition $|M|^2\stackrel{<}{{}_\sim}H^2$, 
for~(\ref{Mpm}) to be valid, requires 
\begin{equation}
 m/\sqrt\varrho\stackrel{<}{{}_\sim}H
\,.
\label{smallm}
\end{equation}
While the latter source in~(\ref{q:dSinvariant}) 
is fully de Sitter invariant, the former source ($\propto \dot H$) 
is obtained based on a heuristic reasoning, according to which the 
quasi de Sitter result can be obtained by taking the appropriate 
time derivative of the Hubble parameter in the de Sitter invariant propagator. 
This conjecture should be tested by performing the corresponding calculations 
with the scalar propagator constructed 
for quasi de Sitter spaces~\cite{JanssenProkopec:2007}.

\section{Decay into baryons}
\label{section:decay}

The above discussion shows that inflationary particle production can directly
result in the generation of a charge asymmetry.
We now present a possible model where this mechanism leads to
a non-zero number of Standard Model baryons.

Let us assume that the neutrino masses are generated by the see-saw mechanism
and that therefore a scalar condensate of VEV $X$, which spontaneously breaks
${\rm U}(1)_{B-L}$, is present. The particle excitations associated with
$X$ are taken to be ultraheavy of order of the Grand Unified scale,
and we do not consider them here. We give the
same ${\rm U}(1)_{B-L}$ charge to $\phi$ as we give to $X$. Then, the
$CP$-violating parameters in the lagrangian~(\ref{L:non_stan:phi})
can be of the origin
\begin{equation}
\omega=h_\omega X^{*2}\,,\qquad \mu^2 = h_\mu X^{*2}\,.
\end{equation}

While $h_\mu$ is a dimensionless parameter, $h_\omega$ is of mass dimension
$-2$, and therefore the $\omega$-term originates from a non-renormalizable
operator. While there is no reason for not allowing such a term at tree-level,
it will generically also be induced by radiative corrections, provided there
are other operators that violate $CP$. For example,
let us assume that there is an input scale $\Lambda\approx M_{\rm GUT}$,
at which the $\omega$-term is absent, and let us add the operators
\begin{equation}
\lambda X \phi^{*2} \phi + \lambda^* X^* \phi^2 \phi^*\,,
\end{equation}
which give rise to cubic interactions
to the lagrangian~(\ref{L:non_stan:phi}). If we assume for simplicity
that $\mu=0$ and $\varrho=1$ at the scale $\Lambda$, the self-energy
acting on the vector $(\begin{array}{cc}\phi & \phi^*\end{array})^T$
reads
\begin{equation}
{\rm i}\Pi(p)=-\frac{\rm i}{8\pi^2}
\left(
\begin{array}{cc}
(\lambda X)^2+(\lambda X)^{*2} & |\lambda X|^2 \\
|\lambda X|^2 & (\lambda X)^2+(\lambda X)^{*2}
\end{array}
\right)
\left(
\log\frac{\Lambda^2}{m^2}+\frac{p^2}{6m^2}
\right)\,,
\end{equation}
where we have expanded in $p^2<4 m^2$ and $m^2 <\Lambda$.
Clearly, $\Pi(p^2)$ induces a radiative correction of the type of
a $\mu$-like mass term and $\partial\Pi(p^2)/\partial p^2$ an
$\omega$-like kinetic term.

However the CP-violating terms are generated, for an efficient charge
production, we have to require $|M^2| \stackrel{<}{{}_\sim}H^2$,
corresponding to some parametric
tuning. We note that for supersymmetric flat directions,
mass terms of Hubble scale appear naturally from supergravity in the presence
of non-vanishing $F$-terms~\cite{Dine:1995kz}. In addition, Hubble scale mass
terms are generated due the SUSY breaking in the expanding
background~\cite{Garbrecht:2006aw}.

We now need to specify how the field $\phi$ couples to Standard Model quarks
and leptons. 
In order to allow for a decay through a tree-level, renormalizable
operator, we postulate a fermion $\widetilde H$, that has the
same gauge quantum numbers as the Standard Model Higgs field $H$.
Assuming that the $B-L$ charge of $\phi$ is now {\it one}, the decay
can be allowed by the renormalizable coupling
\begin{equation}
\label{decay:SM:2}
\phi L \widetilde H+{\rm h.c.}\,.
\end{equation}
Of course, the most obvious example for such a theory is the
Minimal Supersymmetric Standard Model supplemented by a right-handed
neutrino, where $\phi$ should  be identified with the right-handed sneutrino.
This is particularly attractive, since the flatness of the potential $\phi$
arises naturally and is protected against radiative corrections
by supersymmetry. It should be also noted that for a typical scale of
inflation
$H\sim 10^{13}~{\rm GeV}$, the condition
$|M|\stackrel{<}{{}_\sim}H$ is usually satisfied at 
least by one of the heavy Majorana neutrinos~\cite{Buchmuller:2005eh}.

\section{Baryon number of the Universe}
\label{section:BAU}

Now, we estimate the baryon asymmetry which results from
the superhorizon fluctuations. The charge density~(\ref{q:dSinvariant})
has two contributions: first, a term which does not
vanish when $\dot H=0$, which we call contribution~A,
and second from the term $\propto \dot H$, which would vanish
in an exact de~Sitter background and which we denote by contribution~B.
In this section, we calculate the BAU
resulting from both of these terms. The charge and baryon
asymmetries from the
particular contributions are called $q_\phi^{\rm A\,,B}$ and
$n_B^{\rm A\,,B}$, and $q_\phi=q_\phi^{\rm A}+q_\phi^{\rm B}$
as well as $n_B=n_B^{\rm A}+n_B^{\rm B}$.

In order to calculate $n_B/s$, we assume that inflation is followed by a
matter-dominated epoch after which the Universe reheats at a temperature
$T_{\rm R}$ and the entropy density $s$ is produced. Let $a_{\rm end}$ denote
the scale factor at the end of inflation and $a_{\rm R}$ the scale factor at
reheating. We denote the charge density at reheating by $q_\phi^{\rm R}$, and
find for the conserved charge-to-entropy  ratio
\begin{equation}
\label{q:over:s}
\frac{q_\phi^{\rm R}}{s}=\left(\frac{a_{\rm end}}{a_{\rm R}} \right)^3
\frac{q_\phi}{s}
=\frac34 \frac{T_{\rm R}}V q_\phi\,.
\end{equation}
When assuming instant reheating, either by fast perturbative decays or
preheating of the inflaton, one gets the least dilution of the
initially produced charge and
\begin{equation}
\label{instant:reheating}
T_{\rm R}=\left(\frac{30 V}{\pi^2 g_*}\right)^{1/4}
\end{equation}
or $T_{\rm R}/V^{1/4}=0.41$,
where $g_*$ is the number of relativistic degrees of freedom, and we take
here the Standard Model value $g_*=106.75$.

We assume that one unit of $q_\phi$ corresponds to $q_{B-L}$ units
of $(B-L)$-asymmetry, such that $n_{B-L}=q_{B-L} q_\phi^{\rm R}$,
and we note that sphaleron transitions convert the $(B-L)$-asymmetry into
a baryon
number of $n_B=28/79\, n_{B-L}$. Therefore, the BAU is given by
\begin{equation}
\label{n_B:q}
\frac{n_B}s=\frac{21}{79} \frac{T_{\rm R}}V q_{B-L} q_\phi\,,
\end{equation}
which is to be compared to the observed value~\cite{Spergel:2006hy}
\begin{equation}
\label{BAU:observed}
\frac{n_B}{s}\simeq 8.7\times 10^{-11}\,.
\end{equation}

Let us first apply formula~(\ref{n_B:q}) to contribution~A to
the charge asymmetry~(\ref{q:dSinvariant}),
\begin{equation}
q_\phi^{\rm A}=|\omega \mu^2| H \sin \theta_{\rm CP}
\frac{1}{8\pi^2}\frac{1}{\varrho^2 -|\omega|^2}\,.
\end{equation}
Taking account of the relations~(\ref{smallCPviolation}) and~(\ref{smallm}),
it is convenient to express this as $q_\phi^{\rm A}=\tau H^3$,
where $\tau\stackrel{<}{{}_\sim} 1/80$. We find from~(\ref{n_B:q})
\begin{equation}
\frac{n_B^{\rm A}}{s}
=\frac{112}{79}\sqrt{\frac{2 \pi^3}{3}}\frac{T_{\rm R}}{V^{1/4}} 
\left(\frac{V^{1/4}}{m_{\rm Pl}}\right)^3
q_{B-L}\tau
=1.5\times 10^{-9} \frac{T_{\rm R}}{0.41\,V^{1/4}}
\left(\frac{V^{1/4}}{10^{16}{\rm GeV}}\right)^3
q_{B-L}\tau
\,.
\end{equation}

Using the observed value of the BAU~(\ref{BAU:observed}),
taking for definiteness $q_{B-L}=1$ and optimistically assuming
instant reheating~(\ref{instant:reheating}), we find
\begin{equation}
V^{1/4}\approx 3.9 \times 10^{15}{\rm GeV}\tau^{-1/3}
\end{equation}

We see that the baryon asymmetry depends on the magnitude of the scalar
potential during inflation. Within slow-roll inflation
(see {\it e.g.}~\cite{LythRiotto}), it is related to the
tensor-to-scalar ratio $r$ by
\begin{equation}
V\approx(3.3\times 10^{16}{\rm GeV})^4 r\,.
\end{equation}
The present upper bound from WMAP+SDSS, assuming no running of $n_s$,
is $r<0.30$ at
95\% confidence level~\cite{Spergel:2006hy}, which translates to
$V^{1/4}<2.4\times 10^{16}{\rm GeV}$.
Meeting this bound requires $\tau>4.3\times 10^{-3}$.

>From these numbers, it is clear that contribution~A
only yields a sufficient abundance
for large field models of inflation, where $V^{1/4}$ is around
$10^{16} {\rm GeV}$. This bound is met by standard chaotic inflation with
a quadratic potential, where
$V^{1/4}\approx 2.0 \times 10^{16} {\rm GeV}$ at horizon exit and
$V^{1/4}\approx 0.5 \times 10^{16} {\rm GeV}$ at the end of inflation.
This means that contribution~A as a source for the BAU
can be ruled out by future CMB observations, if these constrain
$r<7 \times10^{-2}$, when assuming that $\tau<1/80$.
On the other hand, when assuming
simple chaotic inflation, this source gives an intriguing explanation for the
magnitude of the observed BAU by relating it to the scale of inflation.

Let us now turn to the contribution~B to the charge~(\ref{q:dSinvariant}),
\begin{equation}
\label{qB}
q_\phi^{\rm B}=-|\omega \mu^2| H \sin \theta_{\rm CP}
\frac{3}{2\pi^2}\frac{H^2 \dot H}{m^4 - |\mu|^4}
\,.
\end{equation}
The time-derivative of the Hubble rate is linked to the amplitude of
scalar density perturbations $\sqrt{P_{\cal R}}$ as
\begin{equation}
P_{\cal R}=\frac{2^7\pi}{3m_{\rm Pl}^6}\frac{V^3}{V^{\prime2}}
=-\frac{H^4}{\pi\dot H m_{\rm Pl}^2}\,,
\end{equation}
where we have used the slow-roll approximation $3H\dot \phi=-dV/d\phi$.
The observed value at the scale $0.002{\rm Mpc}^{-1}$ is
$\sqrt{P_{\cal R}}=4.86\times 10^{-5}$.
Introducing
\begin{equation}
\label{xi}
\xi=\frac{|\omega \mu^2|H^2}{m^4-|\mu^4|}\sin \theta_{\rm CP}\,,
\end{equation}
and using~(\ref{q:over:s}), we find
\footnote{Note that in this estimate we have (optimistically) assumed that 
the relevant $\dot H$ in Eq.~(\ref{q:dSinvariant}) 
corresponds to the $\dot H$ at the time of the first Hubble crossing 
for the observed modes of cosmological perturbations, for which $N\simeq 50$.
While this may be the case, it is by no means guaranteed.}
\begin{eqnarray}
\label{nBB}
\frac{n_B^{\rm B}}{s}\!\!\!&=&\!\!\!
\frac{28}{79}q_{B-L}\, \frac{2^{9/2}}{\sqrt{3 \pi}}
\left(\frac{V^{1/4}}{m_{\rm Pl}}\right)^7
\frac{T_{\rm R}}{V^{1/4}}
\frac{\xi}{P_{\cal R}}\\
\!\!\!&=&\!\!\!1.2\times10^{-13}
\frac{T_{\rm R}}{0.41 V^{1/4}}
\left(\frac{V^{1/4}}{10^{16}{\rm GeV}}\right)^7
q_{B-L}\,\xi\,.
\end{eqnarray}
 In this case the relations~(\ref{smallCPviolation}) and~(\ref{smallm}),
which ensure the conservation of $q_\phi$ after inflation and the
validity of our expansions, do not prevent $\xi$ from being arbitrarily large.
However, it takes time until the full charge asymmetry
has built up. For a light scalar field, it is a reasonable assumption
that the term $S_1=3H^4/(8\pi^2M^2)$ in the
propagator~(\ref{Delta:Expanded}) gets replaced by
$S_2=H^2 N_{\rm e}/(4 \pi^2)$, where $N_{\rm e}=Ht$ is the number
of e-folds that have been elapsed since the start of inflation.
When $S_2\simeq|S_1|$, the charge production is saturated and above
results~(\ref{qB}) and~(\ref{nBB}) for $q_\phi^{\rm B}$ and $n_B^{\rm B}$
apply. As long as
$S_2\stackrel{<}{{}_\sim}S_1$, one should replace
$H^2/(m^4-|\mu^4|)\to 2 N_{\rm e}/(3m^2)$ in expression~(\ref{xi}) for $\xi$,
recalling relation~(\ref{smallCPviolation}). For values of
$N_{\rm e}\stackrel{>}{{}_\sim} 60$, just above the minimum number of e-folds
required for a solution of the horizon problem, contribution~B is therefore
also a viable source for baryogenesis, provided $V^{1/4}$ takes large
values of order $10^{16}{\rm GeV}$. For $N_{\rm e}\gg 60$ however,
the contribution B may be much larger than the de Sitter
 invariant contribution A, in which 
case the potential can be smaller and still inflation can 
generate a large enough baryon asymmetry to be consistent with
the observed value.

\section{Conclusions}
\label{section:conclusions}

The CP-violating amplification of vacuum fluctuations during inflation
is a viable mechanism for generation of the BAU. When assuming not much more
than sixty e-folds, a number of conditions needs however
to be met for successful baryogenesis.

We have found that the mechanism only leads to a sufficient baryon abundance
for large scales of inflation, $V^{1/4}\sim 10^{16}{\rm GeV}$. In models
operating at these energy scales, the VEV of the inflaton exceeds the
Planck scale~\cite{Lyth:1996im}.
In the context of local supersymmetry, supergravity
corrections however spoil the slow-roll conditions when the VEVs
exceeds Planck scale. On the other hand, simple chaotic inflation with
a quadratic potential may still be regarded as the simplest possible
implementation of inflation and also deserves interest as it predicts
tensor perturbations, which are observable by near-future CMB observations.
We have also seen that a sufficient asymmetry requires high reheat
temperatures, of the same order as $V^{1/4}$. This again clashes with
local supersymmetry, as it would lead to a cosmological unacceptable
overproduction of gravitinos.

Furthermore, the scalar field needs to decay quickly into fermions
before the asymmetry $q_\phi$ gets altered
in the post-inflationary era. Therefore, we have to assume that
the CP-violating terms are smaller than the CP-conserving ones, as
expressed by relation~(\ref{smallCPviolation}). This also means that
a large Yukawa coupling $f$ is favored. We believe that
a more accurate study of the evolution of the charge asymmetry
after inflation is possible by matching the inflationary solutions
for the scalar modes to those in radiation or matter era and will perform
this in a future publication.

It is also interesting to speculate about eternal
inflation~\cite{Linde:1982ur,Vilenkin:1983xq}. In the simplest
case, the inflaton field is located on a local maximum of a scalar potential.
While it classically tends to roll down the potential, quantum diffusion
countervails such that most domains keep on inflating. Therefore, the
number of e-folds can be huge and potentially a large charge asymmetry
can accumulate. It should however be noted that while the inflaton is
situated at the top of the potential, $\langle \dot H \rangle=0$, and
hence $q_\phi^{\rm B}=0$ as in~(\ref{qB}). However the squared
expectation value $\langle q_\phi^2 \rangle$ does not need to vanish, and
it should be investigated whether it may sufficiently source baryogenesis.

In conclusion, it is not true in general
that inflation dilutes any initial pre-inflationary charge asymmetry. Indeed,
 inflation can be used to generate a nonvanishing baryon asymmetry of the
Universe. If it happened at a high energy scale, 
the scale of inflation will be revealed in the near future
by the observation of tensor modes in the Cosmic Microwave Background.
 In this case the amplified charged quantum fluctuations
 can source a baryon asymmetry of the observed size, 
even under the conservative assumption that 
the total number of e-folds is of the order of sixty. If, on the other hand,
 the scale of inflation happens to be significantly lower, 
a sufficient amount of baryons can still be generated, provided inflation 
lasted much longer than 60 e-folds.

\appendix
\section{Spectrum on superhorizon scales}
\label{appendix:IR}

Starting from the integrand in~(\ref{current:phi:7}),
we define the charge spectrum as 
\begin{eqnarray}
 {\cal P}_{q_\phi}(k,\eta) = \frac{k^3}{2\pi^2} q_{\phi,k}(\eta)
                   &=& \frac{k^3}{2\pi^2}\Bigg\{
-\frac12\frac{|\omega|}{\sqrt{\varrho^2-|\omega|^2}}\partial_t
      \left[
 \left|\Phi_k(\eta)\right|^2_{++} - \left|\Phi_k(\eta)\right|^2_{--}
      \right]
\nonumber\\
 &+& \frac12\frac{\varrho}{\sqrt{\varrho^2-|\omega|^2}}
   \partial_t
     \left[
  \left|\Phi_k(\eta)\right|^2_{+-} -\left|\Phi_k(\eta)\right|^2_{-+}
   \right]
   \Bigg\}\,.
\label{charge:spectrum:2}
\end{eqnarray}
If we are interested in the super-Hubble part of the charge spectrum,
the following small argument expansion of the mode 
function~(\ref{dS:mode function}) is useful
\begin{equation}
 \Phi_k(\eta) \;\;\stackrel{-k\eta\ll 1}{\longrightarrow}\;\;
           \frac{1}{a}\frac{\Gamma(\nu)}{\sqrt{2\pi k}}
                {\rm e}^{i(\pi/2)(\nu-1/2)}
                \left(-\frac{k\eta}{2}\right)^{\frac12 -\nu}
           + {\cal O}(-k\eta)^{\frac32 -\nu}
\,,\qquad  (|M^2|\ll H^2)
\,,
\label{mode function:IR}
\end{equation}  
such that 
\begin{eqnarray}
 |\Phi_k(\eta)|^2 \;\;&\stackrel{-k\eta\ll 1}{\longrightarrow}&\;\;
           \frac{1}{a^2}\frac{\Gamma^2(\nu)}{2\pi k}
                \left(\frac{k}{2Ha}\right)^{1 -2\nu}
           + {\cal O}(-k\eta)^{2 - 2\nu}
\nonumber\\
      &=& \frac{2\Gamma^2(\nu)}{\pi}\frac{H^2}{k^3}
                \left(\frac{k}{2Ha}\right)^{3 -2\nu}
           + {\cal O}(-k\eta)^{2 - 2\nu}
\,,\qquad  (|M^2|\ll H^2)
\,.
\label{mode function:IR:2}
\end{eqnarray}  
When $\nu\rightarrow 3/2$ this reduces to 
\begin{equation}
 |\Phi_k(\eta)|^2 \;\;\stackrel{M\rightarrow 0}{\longrightarrow}\;\;
           \frac{H^2}{2k^3}
           + {\cal O}(-k\eta)^{2 - 2\nu}
\,,\qquad  (k/a\ll H)
\,,
\label{mode function:IR:3}
\end{equation}  
which is the scale invariant spectrum whose time derivative vanishes. 
However, when $M^2\neq 0$, one gets a contribution from the time
 derivative of~(\ref{mode function:IR:2}) 
\begin{equation}
 \partial_t|\Phi_k(\eta)|^2 
   \simeq -\frac{2\Gamma^2(\nu)}{\pi}\frac{H^3}{k^3}
            \left(3-2\nu\right)\left(\frac{k}{2Ha}\right)^{3-2\nu}
\,,\qquad  (k/a\ll H)
\,.
\label{mode function:IR:4}
\end{equation}  
where we used $\partial_t a = Ha$. Since 
\begin{equation} 
  \nu = \frac32 - \frac13\frac{M^2}{H^2} - \frac{1}{27}\frac{M^4}{H^4}
     + {\cal O}(M^6/H^6)\,,
\label{nu:expand}
\end{equation}
Eq.~(\ref{mode function:IR:4}) can be expanded as 
\begin{equation}
 \partial_t|\Phi_k(\eta)|^2 
   \simeq -\frac{1}{3}\frac{M^2H}{k^3}
   \left[
     1-\frac23\frac{M^2}{H^2}\left(\ln\left(\frac{2Ha}{k}\right)
           +\psi(3/2)-\frac{1}{6}\right)
   \right] + {\cal O}\left(\frac{M^6}{H^6}\right)
\,,\quad  (k/a\ll H)
\,.
\label{mode function:IR:5}
\end{equation}  
The charge spectrum~(\ref{charge:spectrum:2})
can now be evaluated on super-Hubble scales to give
\begin{eqnarray}
 {\cal P}_{q_\phi}(k,\eta) &\simeq& \frac{H}{12\pi^2}
\frac{|\omega|}{\sqrt{\varrho^2-|\omega|^2}}
      \left[m_+^2-m_-^2
      \right]
    + {\cal O}(M^4)\,.
\label{charge:spectrum:IR}
\end{eqnarray}
Note that the second line in Eq.~(\ref{current:phi:6}) does not contribute
to the charge at leading order in $M^2$. 
Now making use of Eqs.~(\ref{m2+:2}-\ref{m2-:2}) we finally get
\begin{equation}
 {\cal P}_{q_\phi}(k,\eta) \simeq - \frac{H}{6\pi^2}
\frac{|\omega|}{\varrho^2-|\omega|^2}
|\mu|^2\sin(\theta_{\rm CP})
    + {\cal O}(M^4)
\,.
\label{charge:spectrum:IR:2}
\end{equation}
 Several comments are now in order. Firstly, the leading order contribution
(the one that is linear in $M^2/H^2$) 
to the spectrum of charge is scale invariant, precisely
like the spectrum of matter fluctuations in de Sitter inflation. 
 Furthermore, as expected, the charge production vanishes when either of 
the parameters $\mu$, $\theta_{\rm CP}$ or $\omega$ vanishes, as it should.
This means that the presence of the nonstandard kinetic term
in the lagrangian~(\ref{L:non_stan:phi}) is essential for this 
charge production mechanism. 

 The charge phase space density generated according 
to~(\ref{charge:spectrum:IR:2}) has the important property that
-- just as cosmological perturbations -- it is {\it frozen-in} in the sense 
that it does not decay with the scale factor. 
This is so because the rate of charge dilution due to 
the expansion of the Universe is compensated by the creation of charge
due to the CP violation present in our model~(\ref{L:non_stan:phi}).
As a result, the charge density at the end of inflation is 
independent on the conditions at the beginning of inflation
(provided of course inflation lasts long enough) and 
can be completely expressed in terms of the inflationary 
model parameters, presenting thus an unique opportunity to relate 
inflationary predictions tested by Cosmic Microwave Background and
Large Scale Structure measurements with 
baryon number generation. 

%
%
%
%

\section{Spectrum on subhorizon scales}
\label{appendix:UV}

 Let us now consider charge production on small (sub-Hubble) scales
 generated by our mechanism. For this purpose,
the following asymptotic form of the 
Hankel function is useful,
\begin{eqnarray}
 H_\nu^{(1)}(-k\eta) &=& \sqrt{-\frac{2}{\pi k\eta}}
          {\rm e}^{-ik\eta-i(\pi/2)(\nu+1/2)}
   \Bigg[1-\frac{\Gamma(\nu+3/2)}{\Gamma(\nu-1/2)}\frac{i}{2k\eta}
          -\frac{\Gamma(\nu+5/2)}{\Gamma(\nu-3/2)}\frac{1}{8(k\eta)^2}
\nonumber\\
  &+&\dots 
   +\frac{\Gamma(\nu+n+1/2)}{\Gamma(\nu-n+1/2)}\frac{i^n}{n!(-2k\eta)^n}
    +\, {\cal O}\big((-k\eta)^{-(n+1)}\big)
   \Bigg] 
\,.
\label{Hankel:asymptotic}
\end{eqnarray}
 From this and Eq.~(\ref{dS:mode function}) 
we can get the asymptotic form for the mode function squared (see 8.456
in Ref.~\cite{GradshteynRyzhik:1965}) 
\begin{equation}
 |\Phi_k(\eta)|^2 = \frac{1}{a^2}\frac{1}{2k}
\Bigg[1+\frac12\frac{\Gamma(\nu+3/2)}{\Gamma(\nu-1/2)}
             \left(\frac{Ha}{k}\right)^2
      +\frac38\frac{\Gamma(\nu+5/2)}{\Gamma(\nu-3/2)}
             \left(\frac{Ha}{k}\right)^4
      + {\cal O}\big((aH/k)^6\big)
   \Bigg]
\,,
\label{Phi2:UV}
\end{equation}
such that 
\begin{equation}
 \partial_t|\Phi_k(\eta)|^2 = -\frac{1}{a^2}\frac{H}{k}
           +\frac38a^2\frac{\Gamma(\nu+5/2)}{\Gamma(\nu-3/2)}
             \left(\frac{H}{k}\right)^5
                  + {\cal O}\big((aH/k)^6\big)
\,.
\label{Phi2dot:UV}
\end{equation}
This can be expanded in powers of $M^2/H^2$, 
\begin{eqnarray}
 \partial_t|\Phi_k(\eta)|^2 \simeq -\frac{1}{a^2}\frac{H}{k}
           -\frac34a^2\frac{H^3M^2}{k^5}\left[1-\frac12\frac{M^2}{H^2}\right]
                  + {\cal O}\big(M^2(aH/k)^6\big)
\,.
\label{Phi2dot:UV:2}
\end{eqnarray}
The first (leading) term does not contribute to the 
charge spectrum~(\ref{charge:spectrum:2}), such 
that on sub-Hubble scales the charge generation is suppressed as
\begin{eqnarray}
 {\cal P}_{q_\phi}(k,\eta) 
                   &\simeq& \frac{3}{16\pi^2}\frac{a^2H^3}{k^2}
   \frac{|\omega|}{\sqrt{\varrho^2-|\omega|^2}}
      \left[m_+^2-m_-^2\right]
\,.
\qquad
\label{charge:spectrum:UV}
\end{eqnarray}
 Making use of Eq.~(\ref{m2+:2}-\ref{m2-:2})
we can evaluate the charge spectrum~(\ref{charge:spectrum:UV}), 
\begin{eqnarray}
 {\cal P}_{q_\phi}(k,\eta) 
                   &\simeq& -\frac{3}{8\pi^2}\frac{a^2H^3}{k^2}
 \frac{|\omega|}{\varrho^2-|\omega|^2}|\mu|^2\sin(\theta_{\rm CP})
\,.
\qquad
\label{charge:spectrum:UV:2}
\end{eqnarray}
 An important property of this spectrum is that the charge density
 is ultraviolet finite. Indeed, when the charge 
spectrum~(\ref{charge:spectrum:UV}) is integrated out, 
the dominant contribution comes from the modes $k\sim Ha$, which is 
the most infrared mode for which the asymptotic expansion makes sense. 
We thus conclude that 
$q_\phi|_{\rm UV}\sim |\mu|^2H(|\omega|/\varrho^2)\sin(\theta_{\rm CP})$.
 This then implies that to the leading approximation, one can neglect 
the charge production on sub-Hubble scales in our model.


\end{document}